\documentclass[3p,times,twocolumn]{elsarticle}

\usepackage{ecrc}

\volume{00}

\biboptions{numbers,sort&compress}

\usepackage{amsmath,amssymb,amsfonts}

\def\cL{{\cal L}}

\newcommand{\be}{\begin{equation}}
\newcommand{\ee}{\end{equation}}

\firstpage{1}

\journalname{Nuclear Physics B Proceedings Supplement}

\runauth{}

\jid{nuphbp}

\jnltitlelogo{Nuclear Physics B Proceedings Supplement}

\usepackage{amssymb}

\usepackage[figuresright]{rotating}

\begin{document}

\begin{frontmatter}



\title{Disentangling new physics contributions in  \\[2mm] lepton flavour violating $\tau$ decays}

\author[label1]{Alejandro Celis\fnref{label6}}
\author[label2]{Vincenzo Cirigliano}
\author[label2,label3,label4,label5]{Emilie Passemar}

 \fntext[label6]{Speaker}

\address[label1]{IFIC, Universitat de Val\`encia -- CSIC, Apt. Correus 22085, E-46071 Val\`encia, Spain}
\address[label2]{Theoretical Division, Los Alamos National Laboratory, Los Alamos,  NM 87545, USA}
\address[label3]{ Department of Physics, Indiana University, Bloomington, IN 47405, USA}
\address[label4]{Center for Exploration of Energy and Matter, Indiana University, Bloomington, IN 47403, USA}
\address[label5]{ Theory Center, Thomas Jefferson National Accelerator Facility,
12000 Jefferson Avenue, Newport News, Virginia 23606, USA}

\begin{abstract}
The possibility to discriminate between different operators contributing to lepton flavour violating tau decays is discussed within an effective field theory framework.     Correlations among decay rates in different channels as well as differential distributions in many-body decays are considered.     Recent developments in the determination of the hadronic form factors for $\tau \rightarrow \ell \pi \pi$ ($\ell = e, \mu$) decays are incorporated in the analysis.  The above issues are exemplified by considering a Higgs-like boson with lepton flavour violating couplings.   Implications of the search for lepton flavour violating Higgs decays performed recently by the CMS collaboration are also discussed.     
\end{abstract}

\begin{keyword}
tau decays \sep lepton flavour violation \sep Higgs decays
\end{keyword}

\end{frontmatter}

\section{Introduction}
\label{intro}

The observation of charged lepton flavour violating (LFV) transitions would be a clear indication of physics beyond the Standard Model.  We will be concerned here with LFV $\tau$ decays.      The Belle and BaBar collaborations have stopped collecting data, putting bounds on the branching ratio ($\mathrm{BR}$) of these transitions at the level of $10^{-7}$-$10^{-8}$.     In the future, Belle-II is expected to bring the search for LFV $\tau$ decays to a new level of sensitivity.    The LHCb collaboration could also play an important role for some of these processes, like $\tau \rightarrow 3 \mu$.    

Nothing guarantees that LFV $\tau$ decay rates are within the reach of current and/or future experimental facilities.   It is however well known that many new physics models predict large rates for charged lepton flavour violating transitions, which could be observed experimentally~\cite{Raidal:2008jk}.    In case these transitions are observed in the future it will be crucial to address the following question:

\begin{itemize}
\item How can we discriminate different kinds of new physics in LFV $\tau$ decays?
\end{itemize}
This is discussed in Secs.~\ref{efflag:sec} and \ref{dp:sec} by considering correlations among different LFV $\tau$ decay rates and differential distributions in three-body decays.   The approach taken here is to consider an effective Lagrangian describing $\tau$-$\mu$ LFV transitions.   A series of benchmark scenarios in which only one type of effective operator is relevant will then be defined to frame the discussion.   We do not consider $\tau$-$e$ transitions but the arguments presented here also hold in general for these processes.   

As a specific new physics scenario giving rise to different operators at the low energy scale, we consider the possibility that the recently discovered Higgs boson has sizable LFV couplings.   We show that the pion invariant mass spectrum in $\tau \rightarrow \mu \pi^+ \pi^-$ decays provides a useful handle to unravel the underlying dynamics in this case.   A proper determination of the hadronic form factors near the resonance region, as implemented here,  turns out to be crucial for such purpose~\cite{Daub:2012mu,Celis:2013xja}.    With this scenario in mind,  we also explore the complementarity between the energy and intensity frontiers in probing for LFV phenomena by addressing:
\begin{itemize}
\item   What are the implications of the recent search for LFV Higgs decays performed by the CMS \mbox{collaboration}? 
\end{itemize}
 This is discussed in Sec.~\ref{lfvhiggs:sec}.  We conclude in Sec.~\ref{concl}.  The results presented in this talk have been obtained in Refs.~\cite{Celis:2013xja,Celis:2014asa}. 

\section{Effective Lagrangian}
\label{efflag:sec}

We consider the following effective Lagrangian describing $\tau$-$\mu$ transitions
\begin{align}  \label{tauLAG}
\cL_{\mbox{\scriptsize{eff}}} &\;=\;  \cL_{\mbox{\scriptsize{eff}}}^{(D)} + \cL_{\mbox{\scriptsize{eff}}}^{(\ell q)}  + \cL_{\mbox{\scriptsize{eff}}}^{(G)}   + \cdots \,,
\end{align}
where the effective dipole terms are contained in
\begin{align}  \label{taulag1}
\cL_{\mbox{\scriptsize{eff}}}^{(D)} =&  - \frac{m_{\tau}}{\Lambda^2}  \,   \Bigl\{ \,   \, \left( \mathrm{C_{DR}}  \, \bar \mu \, \sigma^{\rho \nu} \, P_L \, \tau   
+ \mathrm{C_{DL}} \, \bar \mu \, \sigma^{\rho \nu} \, P_R \, \tau \right) F_{\rho \nu}    \nonumber \\
&+ \mathrm{h.c.}  \Bigr\} \,,
\end{align}
and the four-fermion operators are included in
\begin{align}  \label{taulag2} 
\cL_{\mbox{\scriptsize{eff}}}^{(\ell q)} =& - \frac{1}{\Lambda^2}\sum_{q=u,d,s}  \Bigl\{  \left( \mathrm{C^{q}_{VR}}  \, \bar \mu   \gamma^{\rho}  P_R  \tau + \mathrm{C^{q}_{VL}} \, \bar \mu \gamma^{\rho}  P_L  \tau   \right)    \bar q \gamma_{\rho}  q   \nonumber \\ 
&+ \left( \mathrm{C^{q}_{AR}} \,   \bar \mu \, \gamma^{\rho} \,  P_R \, \tau  + \mathrm{C^{q}_{AL}} \,   \bar \mu \, \gamma^{\rho} \, P_L \, \tau \right)   \bar q \, \gamma_{\rho} \gamma_{5} \, q \nonumber \\ 
&+  m_{\tau} m_{q} G_F\,   \left(  \mathrm{C^{q}_{SR}} \, \bar \mu \, P_L \, \tau     + \mathrm{C^{q}_{SL}} \bar \mu  \, P_R \, \tau  \right)   \bar q \,q \nonumber \\  
&+  m_{\tau} m_{q} G_F\,  \left(   \mathrm{C^{q}_{PR}}  \, \bar \mu \, P_L \, \tau   + \mathrm{C^{q}_{PL}}  \,  \bar \mu \, P_R \, \tau  \right)   \bar q \, \gamma_{5}\, q   \nonumber \\
&+ \mathrm{h.c.} \Bigr\}  \,.
\end{align}
The final piece of the effective Lagrangian contains gluonic effective operators
\begin{align}  \label{taulag3}
\cL_{\mbox{\scriptsize{eff}}}^{(G)} &=   - \frac{ m_{\tau} G_F }{\Lambda^2}  \frac{\beta_L}{4 \alpha_s}  \,   \Bigl\{   \, \left(  \mathrm{C_{GR}}  \,  \bar \mu \, P_L \, \tau + \mathrm{C_{GL}} \, \bar \mu \, P_R \,  \tau   \right)   G_{\rho \nu}^{a} G_{a}^{\rho \nu}  \nonumber \\ 
&+\,  \left( \mathrm{C_{ \widetilde{G}R}}  \,  \bar \mu \, P_L \, \tau + \mathrm{C_{ \widetilde{G}L}}  \, \bar \mu \, P_R \,  \tau   \right)  \,    G_{\mu \nu}^{a} \widetilde{G}_{a}^{\mu \nu}
+ \mathrm{h.c.} \Bigr\}  \,.
\end{align}
Here $P_{L,R} = (1 \mp \gamma_5) /2$ are the usual chiral projectors, $G_{\mu \nu}^{a}$ is the gluon field strength tensor and $\widetilde{G}_{\mu \nu}^{a}$ its dual.   Similarly $ F_{\rho \nu} $ denotes the electromagnetic field strength tensor.    The Fermi coupling constant is denoted by $G_F$,  $\beta_L/(4 \alpha_s) = - 9 \alpha_s/(8 \pi)$ and $\Lambda$ represents the high energy scale of the LFV dynamics.  

We have not considered four-lepton effective operators nor tensor operators in Eq.~\eqref{taulag2}.   The discrimination of effective operators in LFV leptonic $\tau$ decays has been discussed in Refs.~\cite{Celis:2014asa,Dassinger:2007ru}.     We will focus here mainly on semileptonic $\tau \rightarrow \ell \pi \pi$ ($\ell = e, \mu$) decays for which significant improvement compared with previous works has been achieved recently, thanks to developments on the relevant hadronic form factors~\cite{Daub:2012mu,Celis:2013xja}.      Semileptonic decays into a pseudoscalar meson $\tau \rightarrow \ell P$ ($P= \pi, \eta^{(\prime)}$) have been discussed within an effective theory language in Ref.~\cite{Celis:2014asa}.
For previous treatments of LFV semileptonic $\tau$ decays, see for example Refs.~\cite{Black:2002wh,Kanemura:2005hr,Herrero:2009tm,Arganda:2008jj,Petrov:2013vka}.

\section{Discriminating effective operators in LFV $\tau$ decays}
\label{dp:sec}

Different new physics scenarios are expected to generate distinctive patterns for the low-energy Wilson coefficients of the effective Lagrangian describing $\tau-\mu$ LFV transitions in Eq.~\eqref{tauLAG}.   Explicit examples of how these effective operators arise after integrating out heavy degrees of freedom have been provided, for example, within the framework of Supersymmetric models~\cite{Brignole:2004ah,Abada:2012cq,Daub:2012mu,Abada:2014kba}, Leptoquark models~\cite{Petrov:2013vka,Wise:2014oea} and left-right symmetric models~\cite{Cirigliano:2004mv}.

To explore the discriminating power to different kinds of operators in LFV $\tau$ decays, we consider a series of benchmark models where only one type of operator is present or relevant.  For simplicity, we also restrict the analysis to the case in which the outgoing muon has a definite chirality.     The benchmark models analyzed here are:

\begin{itemize}

\item {\bf{Gluonic model (Parity-even):}}  In this model only the Parity-even gluonic operator is assumed to be relevant
\be 
\mathrm{C_G}  \equiv  \mathrm{C_{GL}}  \neq 0 \,, \qquad \mathrm{C_{else}} = 0\,.
\ee

\item  {\bf{$Z$-penguin model:}}  Dominance of an effective Z-penguin LFV vertex is assumed. The relative size of the Vector couplings is given in this case by
\begin{align}
\mathrm{C_{Z }}  \equiv  \mathrm{C_{VL}^{u} }       \,, \qquad \mathrm{C_{VL}^{d}}  = (v_d/v_u)     \mathrm{C_{VL}^{u}}     \,,
\end{align}
with
\begin{align}
 v_u &= (1 - \frac{8}{3}  \sin^2 \theta_W )/2 \,,  \,\,  v_d = ( - 1 +  \frac{4}{3}  \sin^2 \theta_W )/2 \,,    
\end{align}
where  $\sin^2 \theta_W \simeq 0.223$ is the weak mixing angle.

\item 

{\bf{Scalar model:}}  The four-fermion scalar operator dominates with a Yukawa-like flavour  structure
\be
\mathrm{ C_S \equiv  C_{SL}^{u}  =  C_{SL}^{d}  = C_{SL}^{s}  }   \neq 0 \,, \qquad \mathrm{C_{else}} = 0     \,.
\ee

\item {\bf{Dipole model:}}  The dipole operator is assumed to dominate.  We set in this scenario
\be 
\mathrm{C_D} \equiv  \mathrm{C_{DL}}   \neq 0 \,, \qquad \mathrm{C_{else}} = 0     \,.
\ee

\end{itemize}

Having defined the different benchmark models, we can now proceed to discuss how correlations between LFV $\tau$ decay rates and differential distributions in three-body decays provide valuable information about the underlying LFV dynamics.   Observables involving polarized $\tau$ decays~\cite{Mannel:2014sba} or searches for $\mu N \rightarrow \tau X$ conversion with high-intensity muon beams~\cite{Sher:2003vi,Kanemura:2004jt} also constitute a complementary handle to unravel the origin of LFV, though they will not be discussed here.

\subsection{Correlations in the decay rates}
\begin{figure}[ht!]
\centering
\includegraphics[width=0.47\textwidth]{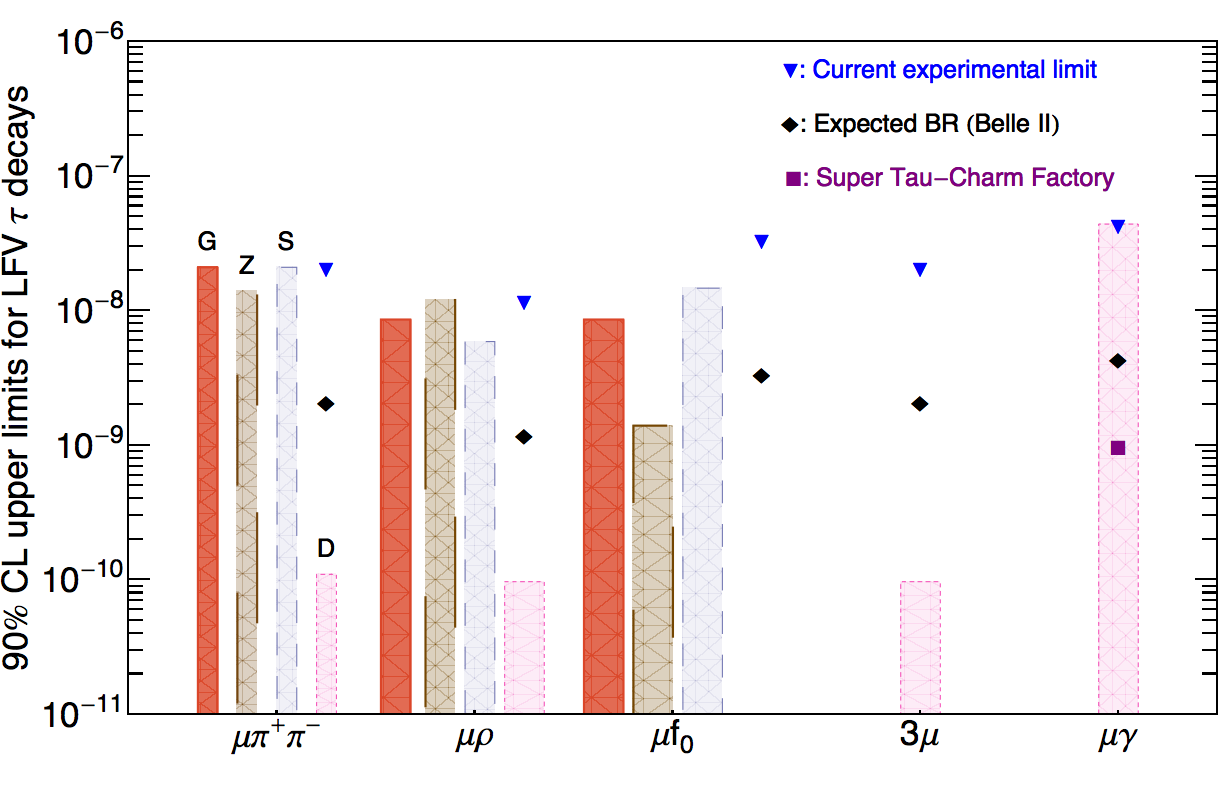}
\caption{\label{fig:prospects} \it \small       Upper bounds on the BR derived from the non-observation of LFV $\tau$ decays for each of the benchmark models considered.       Figure taken from Ref.~\cite{Celis:2014asa}. }
\end{figure}

Assuming that only one type of effective operator dominates, correlations among different LFV $\tau$ decay modes arise.     In case only the effective dipole operator is relevant, what would correspond to our Dipole model, LFV $\tau$ decay rates will be fixed relative to the radiative decay mode $\tau \rightarrow \mu \gamma$.       Fig.~\ref{fig:prospects} shows upper bounds on the BR for different LFV $\tau$ decay modes extracted from the non-observation of these transitions, in each of the benchmark models introduced previously.     Also shown in Fig.~\ref{fig:prospects} are the current experimental upper bounds in each decay channel considered (blue triangles) as well as an estimate of the limits that could be obtained in future factories.  We assume an order of magnitude improvement of the sensitivity at Belle-II (black diamonds).  The bound expected from a Super Tau-Charm Factory on $\mathrm{BR}(\tau \rightarrow \mu \gamma)$ is taken from Ref.~\cite{Bobrov:2012kg} (purple square).  

In the Gluonic model the most sensitive LFV decay mode is $\tau \rightarrow \mu \pi^+ \pi^-$, the BR is given in this case by
\be
\mathrm{BR}(\tau \rightarrow \mu \pi^+ \pi^-) \simeq  0.02 \left( \frac{  \mathrm{C_G} }{\Lambda^2} \right)^2 \, [\mathrm{GeV}^4] \,.
\ee
In the $Z$-penguin model, the most restrictive measurement is that of $\tau \rightarrow \mu \rho$ as can be seen from Fig.~\ref{fig:prospects}.  The BR for $\tau \rightarrow \mu \pi^+ \pi^-$ is given by
\be
\mathrm{BR}(\tau \rightarrow \mu \pi^+ \pi^-) \simeq 1.4 \times 10^{10} \left( \frac{ \mathrm{C_{Z}}    }{\Lambda^2} \right)^2 [\mathrm{GeV}^4]  \,.
\ee
For the Scalar model we obtain
\be
\mathrm{BR}(\tau \rightarrow \mu \pi^+ \pi^-) \simeq 1.9 \times 10^{-3} \left( \frac{\mathrm{C_S} }{\Lambda^2} \right)^2 \, [\mathrm{GeV}^4] \,.
\ee
The strongest constraint is coming in this case from the present limits on $\tau \rightarrow \mu \pi^+ \pi^- $.   Finally, for the Dipole model the most sensitive channel is naturally $\tau \rightarrow \mu \gamma$ for which:
\be
\mathrm{BR}(\tau \rightarrow \mu \gamma) \simeq 6.2 \times 10^{11} \left( \frac{ \mathrm{C_D} }{\Lambda^2} \right)^2 \, [\mathrm{GeV}^4] \,.
\ee
The message of this subsection is clear.  If LFV $\tau$ decays are observed in the future, a combination of measured LFV $\tau$ decay rates together with upper bounds on other non-observed LFV $\tau$ decay channels will provide the main tool to discriminate different types of new physics.   For a more complete discussion see Ref.~\cite{Celis:2014asa} and references therein.

\subsection{Differential distributions in three-body decays}
Assuming LFV $\tau$ decays are observed, we would like to gain as much information as possible about the underlying new physics.  Besides the information provided by a comparison of different LFV $\tau$ decay channels, discussed in the previous subsection, a natural step forward would be to exploit differential distributions in LFV three-body decays.   Of course, we assume here that such transitions would be observed.  

The discrimination of different effective operators in three-body leptonic decays like $\tau \rightarrow 3 \mu$ has been discussed in detail in Refs.~\cite{Celis:2014asa,Dassinger:2007ru}.  In this case a Dalitz plot analysis could be used to determine the dominant operators.   The main obstacle for such analysis would be the low number of collected events, triggering the interest in observables involving polarized $\tau$ decays which might be more useful having a small sample of events~\cite{Mannel:2014sba}.

Semileptonic decays $\tau \rightarrow \ell \pi^+ \pi^-$ also contain information about the underlying new physics in the pion invariant mass spectrum.  Counting the number of events as a function of the di-pion invariant mass can be used to disentangle different effective operators.    The di-pion invariant mass $s=(p_{\pi^+} + p_{\pi^-} )^2$ is kinematically limited in these decays to 
\be
 4 m_{\pi}^2  \leq  s  \leq (  m_{\tau} - m_{\mu} )^2 \,.
\ee
This raises an important issue.   The pion invariant mass can reach values $\sqrt{s} \sim 1$~GeV which are well above the regime of validity of Chiral Perturbation Theory (ChPT) as a low energy effective theory of QCD.   In other words, a determination of the hadronic form factors entering in $\tau \rightarrow \ell \pi \pi$ decays  based on ChPT alone is not reliable in all the accessible kinematical range.  Claims that large deviations from the ChPT predictions are not to be expected in the chiral limit~\cite{Petrov:2013vka}, miss the point that even in this limit, ChPT is inadequate to describe the hadronic dynamics for large invariant masses of the pion pair.     A proper estimation of the decay rate and the pion invariant mass spectrum can be obtained by a combination of ChPT and dispersive techniques as done in Refs.~\cite{Daub:2012mu,Celis:2013xja,Celis:2014asa}.   

\begin{figure}[ht!]
\centering
\includegraphics[width=0.47\textwidth]{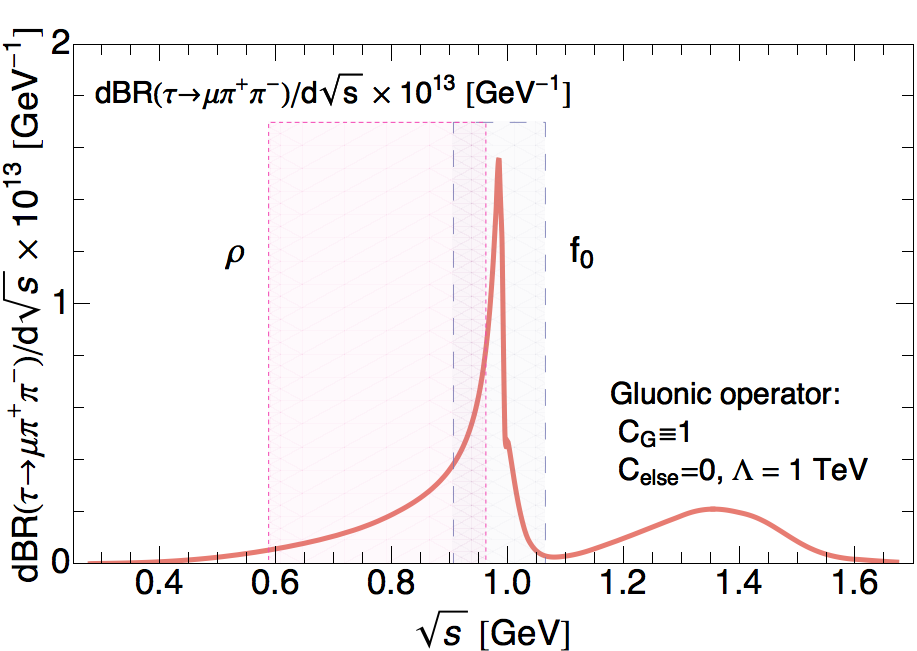}
\caption{\label{fig:Diff} \it \small   Pion invariant mass spectrum in $\tau \rightarrow \mu \pi^+ \pi^-$ decays in the Gluonic model.     Figure taken from Ref.~\cite{Celis:2014asa}.  }
\end{figure}

\begin{figure}[ht!]
\centering
\includegraphics[width=0.47\textwidth]{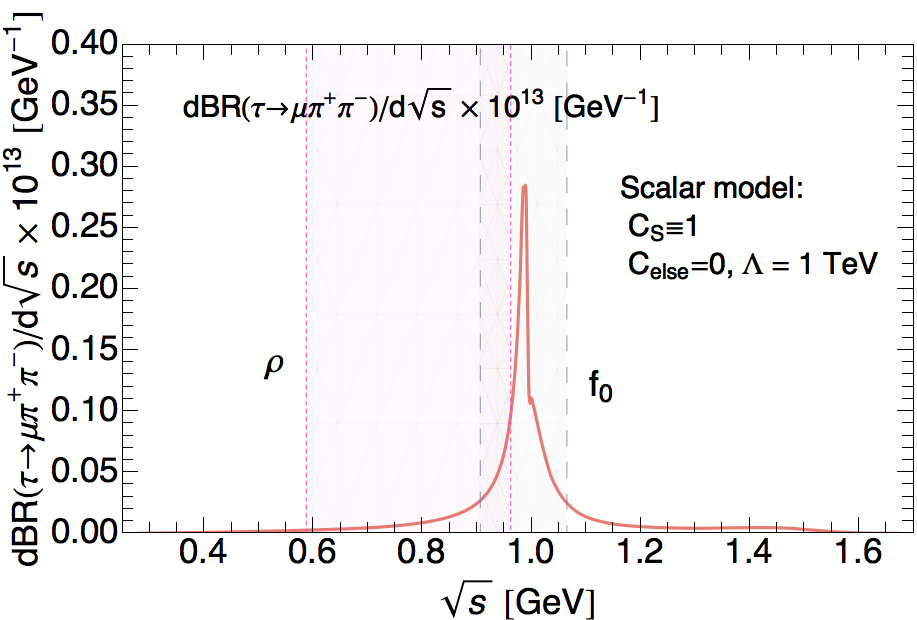}
\caption{\label{fig:DiffS} \it \small   Pion invariant mass spectrum in $\tau \rightarrow \mu \pi^+ \pi^-$ decays in the Scalar model.   Figure taken from Ref.~\cite{Celis:2014asa}.  }
\end{figure}

\begin{figure}[ht!]
\centering
\includegraphics[width=0.45\textwidth]{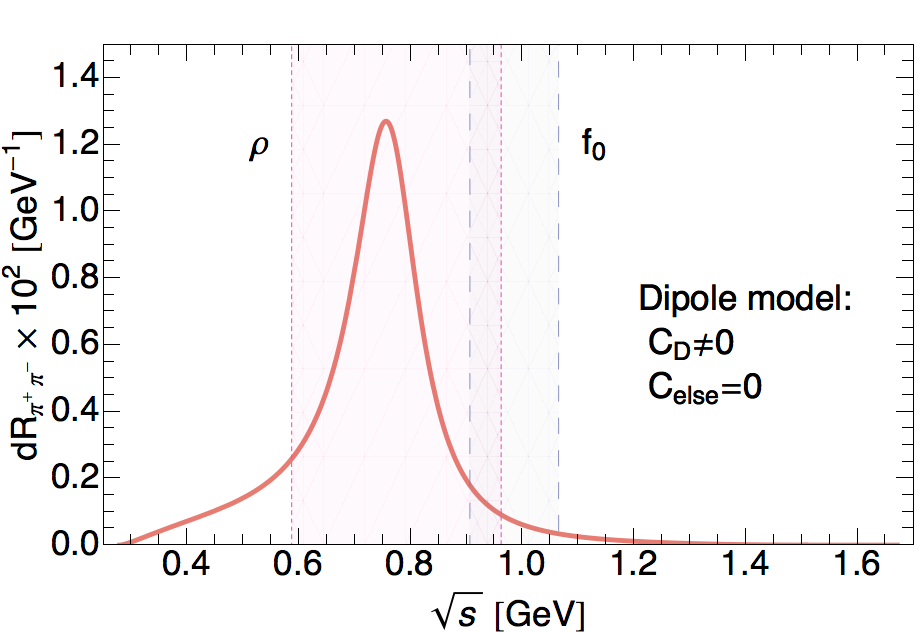}
\caption{\label{fig:DiffD} \it \small   The ratio $dR_{\pi^+\pi^-}$ showing the pion invariant mass spectrum in $\tau \rightarrow \mu \pi^+ \pi^-$ decays within the Dipole model. Figure taken from Ref.~\cite{Celis:2014asa}.  }
\end{figure}

Figs.~\ref{fig:Diff}, \ref{fig:DiffS} and \ref{fig:DiffD} show the pion invariant mass spectrum in $\tau \rightarrow \mu \pi^+ \pi^-$ decays for the Gluonic, Scalar and Dipole models respectively. The pink (short-dashed) and gray (long-dashed) bands in these figures denote the experimental cuts on the pion invariant mass used to search for $\tau \rightarrow \mu \rho$ and $\tau \rightarrow \mu f_0$ respectively~\cite{Miyazaki:2011xe,Miyazaki:2008mw}.        For the Gluonic and Scalar models we simply show the differential BR as a function of $\sqrt{s}$, a peak around the $f_0$ hadronic resonance is clearly observed in both cases.    In Fig.~\ref{fig:DiffD} we have plotted for convenience the ratio 
\be
dR_{\pi^+\pi^-} \equiv \frac{d\Gamma(\tau \rightarrow \mu \pi^+ \pi^-  )/d\sqrt{s}}{ \Gamma(\tau \rightarrow \mu \gamma) } \,.
\ee
Note that all the dependence on $\mathrm{C_D}/\Lambda^2$ cancels in $dR_{\pi^+\pi^-}$.     In the Dipole model, the pion invariant mass spectrum is determined by the pion vector form factor and peaks around the $\rho$ mass $(m_{\rho} \sim 770$~MeV).

\section{LFV Higgs couplings and semileptonic $\tau$ decays}
\label{lfvhiggs:sec}
\begin{figure}[ht!]
\centering
\includegraphics[width=0.2\textwidth]{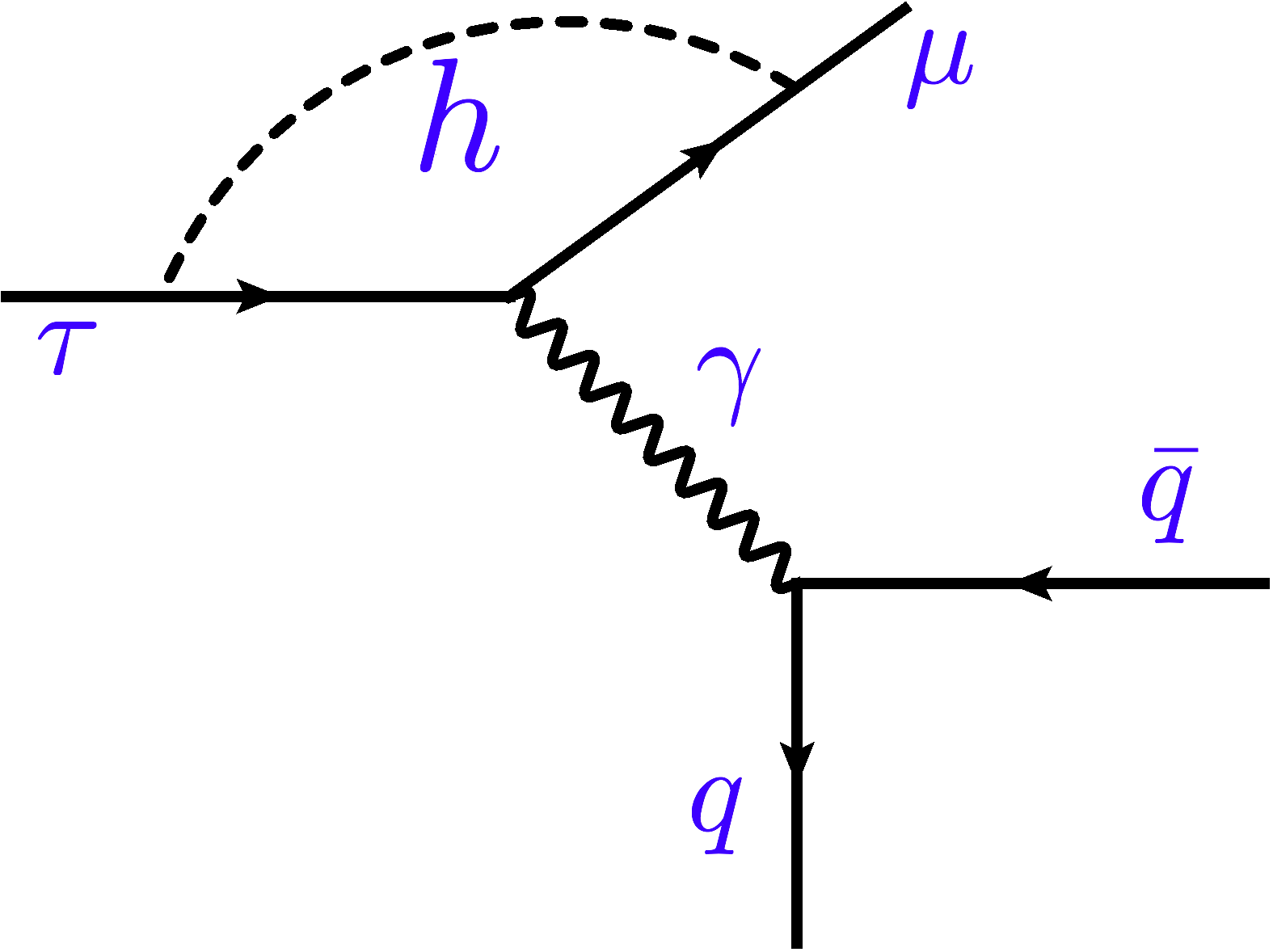}
\caption{\label{fig:eff1} \it \small   Photon mediated contribution to $\tau \rightarrow \mu \pi\pi$.}
\end{figure}

\begin{figure}[ht!]
\centering
\includegraphics[width=0.2\textwidth]{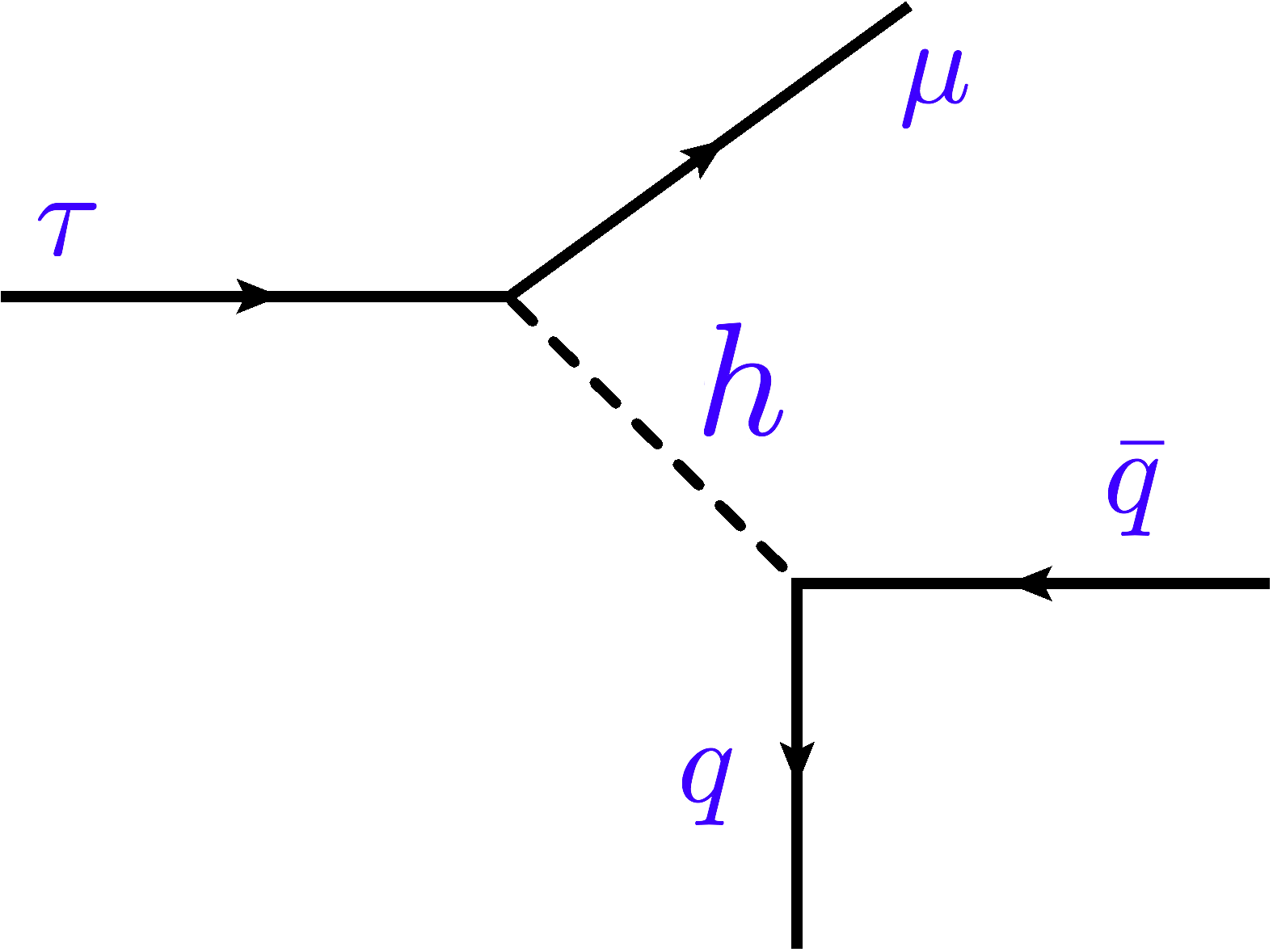}
~~~~
\includegraphics[width=0.18\textwidth]{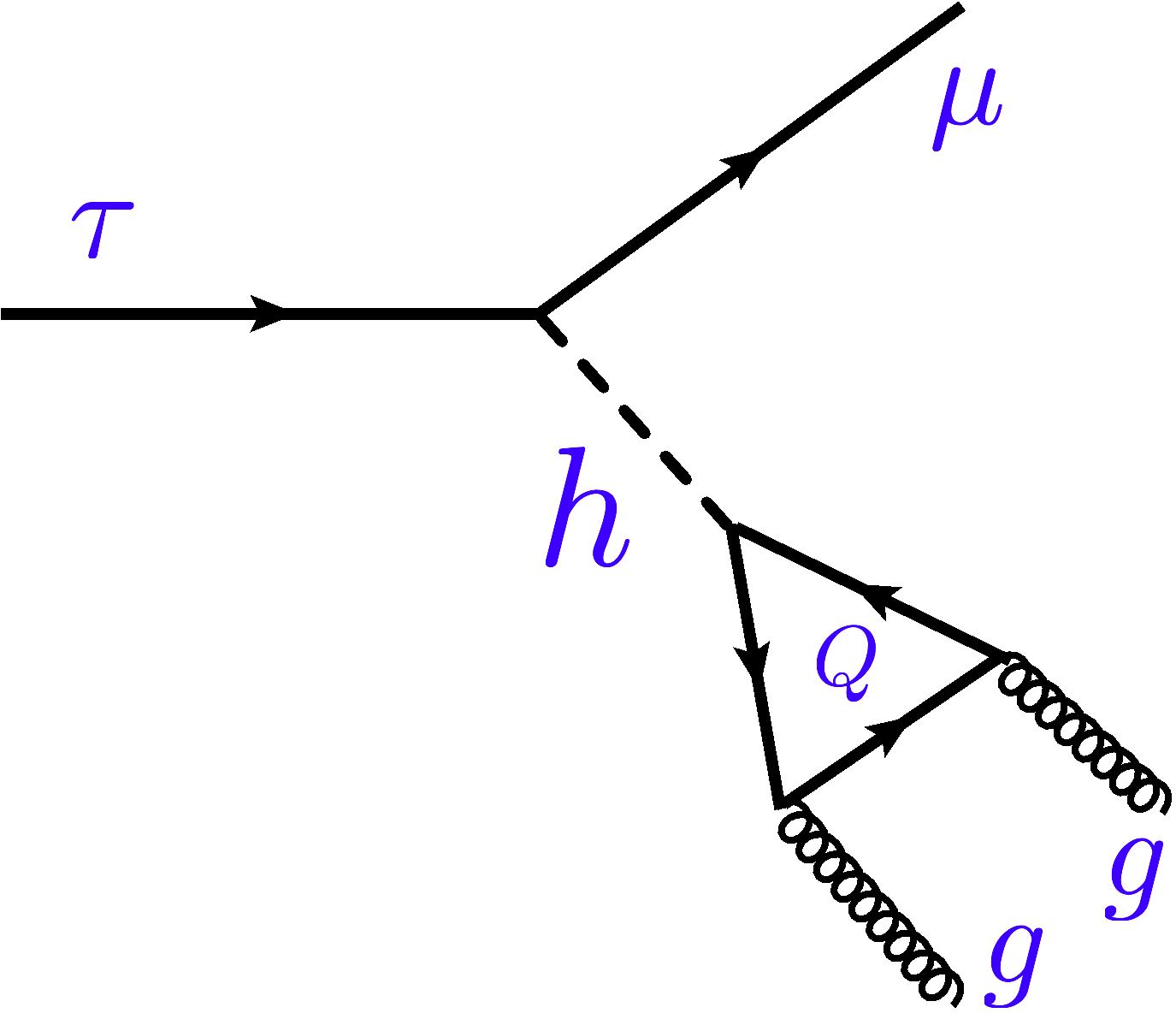} 
\caption{\label{fig:eff2} \it \small   Higgs mediated contribution to $\tau \rightarrow \mu \pi\pi$.   }
\end{figure}

We consider in this section the possibility that the recently discovered Higgs boson with mass around 125~GeV has sizable flavour violating coupling to leptons~\cite{Pilaftsis:1992st,DiazCruz:1999xe,Han:2000jz,Assamagan:2002kf,Sher:2002ew,Paradisi:2005tk,Goudelis:2011un,Blankenburg:2012ex,Harnik:2012pb,Davidson:2012ds,Crivellin:2013wna,Dery:2014kxa,Campos:2014zaa},  
\begin{align} \label{lagGEN}
\mathcal{L}_{Y} \supset -  h \,\left\{  \, Y^{h}_{\tau \mu}  \left( \bar \tau_L \, \mu_{R}  \right)  +  Y^{h}_{\mu \tau}   \left( \bar \mu_L \, \tau_{R}  \right) + \mathrm{h.c.} \right\} \,.
\end{align}
Such couplings could arise from an extended Higgs sector or from effective operators of dimension six encoding details of the high energy dynamics, see Ref.~\cite{Celis:2013xja} and references therein.   Effective dipole operators appear at low energy via Higgs mediated loop diagrams as that shown in Fig.~\ref{fig:eff1}.  Scalar and gluonic operators also appear due to the diagrams shown in Fig.~\ref{fig:eff2}.  

To describe Higgs mediated $\tau \rightarrow \ell \pi \pi$ decays it is crucial to take into account:
\begin{itemize}
\item The Higgs coupling with strange quarks and the effective coupling to gluons induced by heavy quarks. 
\item  A proper determination of the hadronic form factors in all the kinematical range, specially in the resonance region.    
\end{itemize}
All these points have been considered for the first time in Ref.~\cite{Celis:2013xja}.   Effective gluonic interactions induced by heavy quarks were not being included in previous analyses of Higgs mediated $\tau \rightarrow \ell \pi \pi$ decays.  These effects are known to play an important role in the context of $\mu$-$e$ conversion in Nuclei~\cite{Shifman:1978zn,Crivellin:2014cta} and very light Higgs decays~\cite{Donoghue:1990xh}.     Similarly, effective gluonic interactions play a crucial role for $\tau \rightarrow \ell \pi \pi$ decays~\cite{Celis:2013xja}.

\begin{figure}[ht!]
\centering
\includegraphics[width=0.45\textwidth]{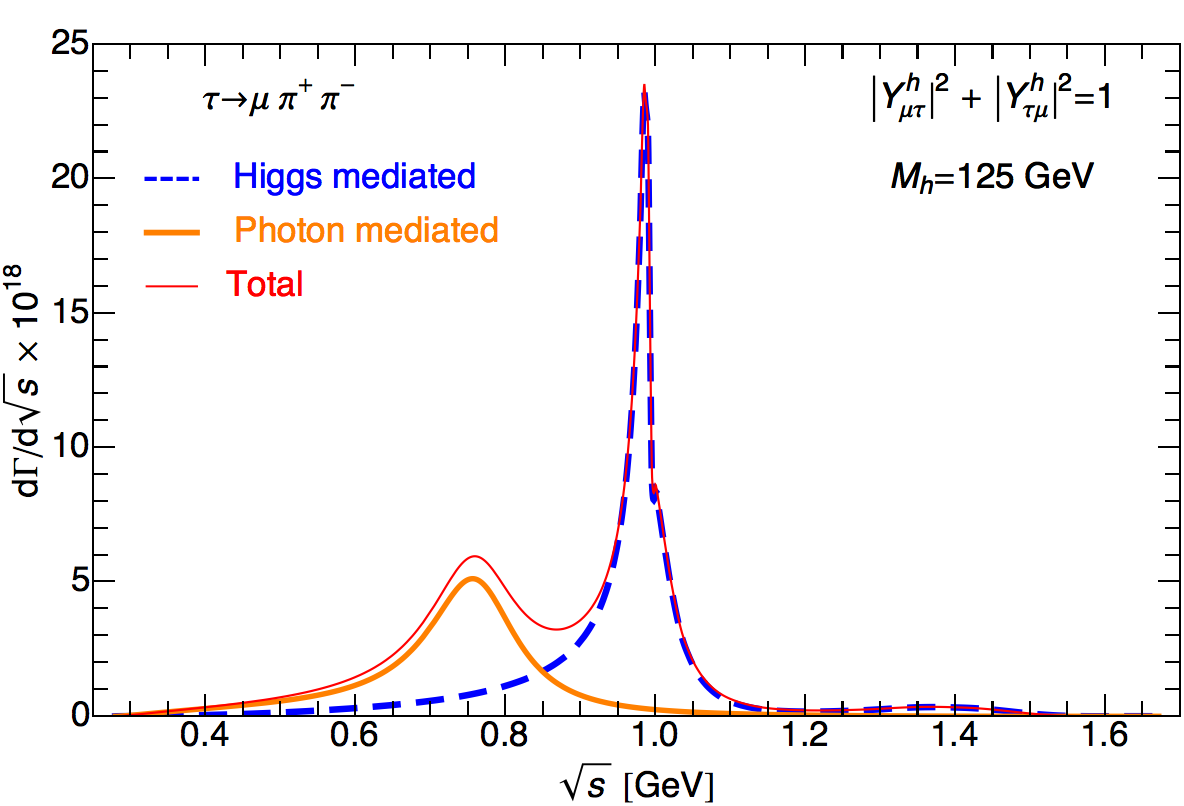}
\caption{\label{fig:DiffH} \it \small   Pion invariant mass spectrum in $\tau \rightarrow \mu \pi^+ \pi^- $ decays mediated by a Higgs boson with LFV couplings.   The Higgs mass is taken to be $M_h = 125$~GeV, the LFV Higgs couplings are fixed to $   |Y_{\mu \tau }^{h}|^2 +  |Y_{\tau \mu }^{h}|^2~=~1  $ and its diagonal couplings are assumed to be SM-like.     Figure taken from Ref.~\cite{Celis:2013xja}. }
\end{figure}

Fig.~\ref{fig:DiffH} illustrates the role of the pion invariant mass spectrum in unraveling the origin of LFV in $\tau$ decays.   In this figure we show the differential decay width for $\tau \rightarrow \mu \pi^+ \pi^-$ as a function of $\sqrt{s}$.   The Higgs mediated contribution is shown in blue (dashed) while the photon mediated one is shown in orange (continuous-thick).  The total differential rate is shown in red (continuous-thin).   We have fixed in this case the Higgs mass to $125$~GeV, the LFV couplings to $   |Y_{\mu \tau }^{h}|^2 +  |Y_{\tau \mu }^{h}|^2~=~1$ and the Higgs diagonal couplings have been taken to be SM-like.     The photon mediated contribution includes two-loop diagrams of the Barr-Zee type calculated in Ref.~\cite{Chang:1993kw} and recently discussed in Ref.~\cite{Harnik:2012pb}.       If these decays are observed, the pion invariant mass spectrum would allow to disentangle the scalar and photon mediated contributions to this process.

\begin{table}[t!]\begin{center}
\vspace{0.2cm}
\begin{tabular}{|c|c|c|c|c|}
\hline
Process &  $(\mathrm{BR \times 10^{8}} )~90\%$ CL&   $ \rule{0cm}{0.5cm} \sqrt{ |Y^{h}_{\mu \tau}|^2 + |Y^{h}_{\tau \mu}|^2  }$   \\ 
\hline  \hline
$\tau \rightarrow \mu \gamma$ &  $<4.4$~~\cite{Aubert:2009ag} & $<0.016$   \\  \hline
$\tau \rightarrow \mu \mu \mu$ &  $<2.1$~~\cite{Hayasaka:2010np}  & $<0.24$    \\  \hline
$\tau \rightarrow \mu \pi^+ \pi^-$ &  $<2.1$~~\cite{Miyazaki:2013yaa}   & $<0.13$   \\  \hline
$\tau \rightarrow \mu \rho$ &  $< 1.2$~~\cite{Miyazaki:2011xe}   &    $<0.13$    \\  \hline 
$\tau \rightarrow \mu f_0$ &  $< 3.4$~~\cite{Miyazaki:2008mw}   &    $<0.26$    \\  
\hline
\end{tabular}
\caption{\it \small Current experimental limits on different LFV $\tau$ decays and bounds extracted on possible LFV Higgs couplings.  The Higgs mass is fixed at $125$~GeV and the Higgs diagonal couplings are taken to be SM-like~\cite{Celis:2013xja}.   }
\label{tab:CPeven}
\end{center}\end{table}

Current bounds on LFV Higgs couplings from $\tau \rightarrow \mu$ decays are summarized in Table~\ref{tab:CPeven}.    Certainly, the strongest bound is coming from the radiative mode $\tau \rightarrow \mu \gamma$.   Note however that the decay rate for this process receives very large contributions from two-loop diagrams and is therefore very sensitive to details of the UV completion of the theory.  Additional particles with LFV couplings can enter at the same level than the Higgs boson causing interfering contributions.    The same happens for $\tau \rightarrow 3 \mu$ for which the same kind of two-loop diagrams are present, by attaching the photon to a $\mu^+ \mu^-$ pair.   In this sense, we consider the bound coming from $\tau \rightarrow \mu \pi^+ \pi^-$ as a complementary handle that should not to be neglected in phenomenological discussions.   The fact that the extracted limit on the LFV couplings is the same from the $\mu \pi^+ \pi^-$ and $\mu \rho$ modes is a mere coincidence.

\begin{figure}[ht!]
\centering
\includegraphics[width=0.48\textwidth]{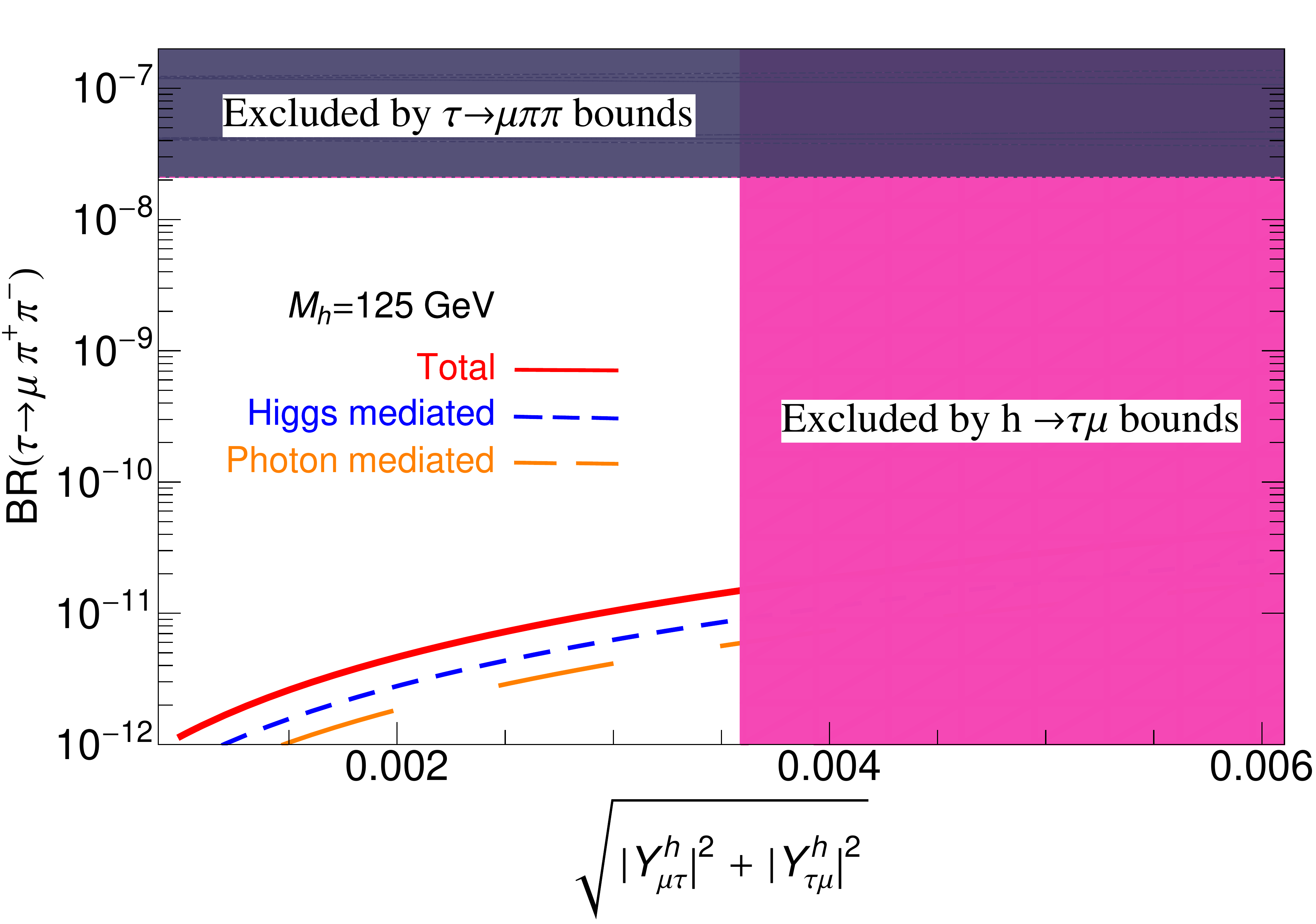}
\caption{\label{fig:DiffH2} \it \small  Constraints on LFV Higgs couplings from $\tau \rightarrow \mu \pi^+ \pi^-$ decays compared with bounds obtained from the search of LFV Higgs decays~\cite{CMS:2014hha,Jan}.      Higgs diagonal couplings are assumed to be SM-like as a benchmark, see text for details. Figure adapted from Ref.~\cite{Celis:2013xja}.  }
\end{figure}

After the Higgs discovery, phenomenological studies analyzed the possibility to observe LFV Higgs decays at the LHC~\cite{Davidson:2012ds,Harnik:2012pb,Bressler:2014jta}. Recently, the CMS collaboration has performed a search for LFV Higgs decays $h \rightarrow \tau \mu$~\cite{CMS:2014hha,Jan}, setting the following upper bound at 95\%~\text{CL.},
\be 
\mathrm{BR}(h \rightarrow \tau \mu) \leq 1.57\% \,, \qquad    \sqrt{   |Y_{\mu \tau}^{h}|^2  + |Y_{\tau \mu}^{h}|^2   } \leq 0.0036 \,.
\ee
It is important to stress a couple of points when interpreting this bound as limits on the LFV Higgs couplings:
\begin{itemize}

\item All the diagonal Higgs couplings are taken to be SM-like, meaning that $g_{hVV} = (g_{hVV})_{\mathrm{SM}} $ and $g_{h\bar f f} = (g_{h\bar f f})_{\mathrm{SM}}$.  Here $VV = (W^+W^-, ZZ)$ and $f$ stands for any of the SM fermions.  The only exotic Higgs couplings are those given in Eq.~\eqref{lagGEN}.

\item The decay $h \rightarrow \tau \mu$ is assumed to be the only relevant non-SM decay channel of the Higgs boson. That is, the total width is $\Gamma(h) = \Gamma(h)_{\mathrm{SM}} + \Gamma(h \rightarrow \tau \mu)$ with
\begin{align}
\Gamma(h \rightarrow \tau \mu) \equiv& \;   \Gamma(h \rightarrow \tau^+ \mu^- + \mu^- \tau^+ )   \nonumber \\
=& \; \dfrac{ M_h }{  8 \pi}   \left(     |  Y_{\mu \tau}^{h} |^2 + |   Y_{ \tau \mu}^{h}|^2 \right)   \,.
\end{align}
\end{itemize}
When working with a specific new physics model none of these assumptions will remain valid in general.   In this sense, the bound on  $\sqrt{   |Y_{\mu \tau}^{h}|^2  + |Y_{\tau \mu}^{h}|^2   }$ presented previously should be interpreted with care.    To have an idea of the complementarity between the search for LFV Higgs decays and the search for LFV $\tau$ decays, we translate the bounds on $\mathrm{BR}(h \rightarrow \tau \mu)$ into limits over the low energy transitions:
\begin{align} \label{eq2}
\mathrm{BR}( \tau \rightarrow \mu \gamma ) &\leq 2.3 \times 10^{-9} \,,\nonumber \\
\mathrm{BR}( \tau \rightarrow 3 \mu ) &\leq 4.7 \times 10^{-12} \,, \nonumber \\
\mathrm{BR}( \tau \rightarrow \mu \pi^+ \pi^- ) &\leq 1.5 \times 10^{-11} \,.
\end{align}
Where again, we have kept the diagonal Higgs couplings to their SM-value.   Two-loop contributions of the Barr-Zee type have been taken into account.  See discussions in Ref.~\cite{Celis:2013xja} for more details on the evaluation of the decay rate for these processes.    Fig.~\ref{fig:DiffH2} shows a comparison between the limits on LFV Higgs couplings extracted from $h\rightarrow \tau \mu$ with those from $\tau \rightarrow \mu \pi^+ \pi^-$.   Predictions for $\mathrm{BR}( \tau \rightarrow \mu \pi^+ \pi^- )$ are shown with the same conventions than in Fig.~\ref{fig:DiffH}: Higgs mediated (blue), photon mediated (orange) and total rate (red).   The main message is that Higgs mediated LFV $\tau \rightarrow \mu$ transitions are strongly constrained by the CMS limit on $\mathrm{BR}(h \rightarrow \tau \mu)$~\cite{CMS:2014hha,Jan}.   The observation of LFV Higgs decays at the LHC remains an interesting possibility allowed by low energy constraints.   If nonzero rates for these decay modes are measured at some point, the search for CP violating effects would also reveal features of the underlying LFV dynamics~\cite{Kopp:2014rva}.

\section{Conclusions}
\label{concl}

In case LFV $\tau$ decays are observed in the future, correlations between the decay rates will provide the main handle for the determination of the underlying dynamics.    Additionally, differential distributions in three-body decays like $\tau \rightarrow 3 \mu$ or $\tau \rightarrow \mu \pi \pi$ provide a complementary handle to discriminate different kinds of new physics.     Finally, with the discovery of a Higgs boson around $125$~GeV, a new window opens in the search for LFV phenomena through Higgs decays.   Low energy constraints on LFV Higgs couplings still allow for sizable effects to be observed at the LHC or at a future Higgs factory.

\section*{Acknowledgments}
The work of A.C. has been supported in part by the Spanish Government and ERDF funds from the EU Commission [Grants FPA2011-23778 and CSD2007-00042 (Consolider Project CPAN)].    The work of V.C. and E.P. is supported by the DOE   Office of Science, Nuclear Physics program. 

\end{document}